\documentstyle[12pt]{article}
\newfont {\xx} {cmti10}

\begin{document}
\def\smp{Standard Model. }
\def\sm{Standard Model }
\def\be{\begin{equation}}
\def\l{\label}
\def\r{\ref}
\def\ee{\end{equation}}
\def\bea{\begin{eqnarray}}
\def\eea{\end{eqnarray}}
\def\nn{\nonumber \\}
\def \R{(\frac{ \alpha_i(0)}{ \alpha_i(t)})}
\def \RI{(\frac{\alpha_i(t)}{\alpha_i(0)})}
\def \R3{(\frac{\alpha_3(0)}{\alpha_3(t)})}
\def \RI3{(\frac{\alpha_3(t)}{\alpha_3(0)})}
\def \EP{($\frac{\alpha(t)}{\alpha(0)})^B$ }
\def \EPP{($\frac{\alpha(M_X)}{\alpha(M_C)})^B$ }
\def\FPR{($\frac{\tilde\alpha(0)}{Y_t(0)})/(\frac{\tilde\alpha}
{Y_t})^*_{MSSM}$}
\def \FPRX{($\frac{\tilde\alpha(M_X)}{Y_t(M_X)})/
(\frac{\tilde\alpha}{Y_t})^*_{MSSM}$}
\def
\FPRP{($\frac{\tilde\alpha(M_P)}{Y_t(M_P)})/
(\frac{\tilde\alpha}{Y_t})^*_{MSSM}$}

\title{Infra-red fixed points revisited.}

\author{Marco Lanzagorta$^a$,   \\Graham G. Ross$^b$\thanks{SERC
Senior
Fellow, on leave from $^a$},\\
\\$^a $Department of Physics,
Theoretical Physics,\\
University of Oxford,
1 Keble Road,
Oxford OX1 3NP\\ \\ $^b$ Theory Division, CERN, CH-1211, Geneva, Switzerland}

\date{}
\maketitle
\abstract{We reconsider how Yukawa couplings may be determined in
terms of a gauge coupling through the infra-red fixed point structure
paying particular regard to the rate of approach to the fixed point.
Using this we determine whether the fixed point structure of an
underlying unified theory may play a significant r$\hat{o}$le in
fixing the couplings at the gauge unification scale. We argue that,
particularly in the case of compactified theories, this is likely to
be the case and illustrate this by a consideration of
phenomenologically interesting theories. We discuss in particular
what the infra-red
fixed point structure implies for the top quark mass.}

\vspace{-15cm}\hspace{11cm} CERN-TH.7522/94

\hspace{11cm} OUTP 9437P

\vspace{15cm}

\newpage

\section{Introduction}
The idea that there may be a stage of unification beyond the standard
model has led to the realisation of the importance of radiative
corrections in determining the dimension $\le 4$ terms in the
lagrangian. This, together with an assumption about unification of
couplings, has led to the successful prediction of the ratio of gauge
couplings in the minimal supersymmetric  extension of the \sm with a
gauge unification scale $M_X\approx 10^{16}GeV$. The same ideas
applied to the soft SUSY breaking terms leads to a convincing
description of the origin of electroweak breaking. In this paper we
consider again whether radiative corrections can also determine the
Yukawa couplings and hence the masses and mixing angles of the theory.
The question is made more timely by the evidence for a top quark with
a mass of $O(174GeV)$ for a large mass is characteristic of the
expectations following from the infra-red fixed point structure of the
\sm
and its extensions.

We start by reviewing the fixed point structure found in
the \smp Keeping only the top Yukawa coupling, $h_t$,  and the
gauge couplings, $g_{3,2,1}$, we have

\be
8\pi^2\frac{d ln(h_t)}{d t}=8g_3^2+\frac{3}{4}(3g_2^2+g_1^2)
+\frac{2}{3}g_1^2-\frac{9}{2}h_t^2
\l {eq:1}
\ee
where $t=ln (\frac{\mu_0^2}{\mu^2})$ and $\mu_0$ and $\mu$ are the
initial
 and final scales at which the couplings are determined.

If one ignores the smaller gauge couplings $g_2$ and $g_1$ the
equation has a fixed point structure which relates the top Yukawa
coupling to the QCD
coupling $g_i$.  We have

\be
8\pi^2\frac{d ln(\frac{h_t}{g_3})}{d t}=g_3^2-\frac{9}{2}h_t^2
\l {eq:2}
\ee
giving the infra-red stable fixed point value \cite{pr}

\be
(h_t^2)^*=\frac{2}{9}g^2_3
\l {eq:fp}
\ee

However, as stressed by Chris Hill \cite{hill}, this fixed point value is not
reached for large initial values of the top quark coupling because the
range in t as $\mu$ varies between  between the Planck scale and
the electroweak scale is too
small to cause the trajectories closely to approach the fixed point.
Rather Hill showed a ``Quasi fixed point'' governs the value of $h_t$
for large initial values of $h_t$. To exhibit
this it is useful to provide a complete analytic solution to eq(\r{eq:2}).

\section{Analytic solution of the renormalisation group equations}
\label{sec:2}
We first solve the general form of the renormalisation group equations
for the case of a single dominant Yukawa coupling and several gauge couplings

\bea
\frac{d g_i^2}{d t}&=&-\frac{b_i g_i^4}{(4\pi)^2} \nn
\frac{d Y_t}{dt}&=&Y_t(\sum_i r_i \tilde{\alpha_i}- s Y_t)
\l {eq:rg}
\eea
where

\bea
\tilde{\alpha_i}&=&\frac{g_i^2}{(4\pi)^2} \nn
Y_t&=&\frac{h_t^2}{(4\pi)^2} \nn
\beta_i&=&\tilde{\alpha}_i(0) b_i
\l {eq:def}
\eea

The solution to these equations is \cite{il}

\bea
\tilde\alpha_i(t)&=&\frac{\tilde\alpha_i(0)}{(1+\beta_i t)} \nn
Y_t(t)&=&\frac{Y_t(0)E_1(t)}{1+sY_t(0)F_1(t)}
\l {eq:rgs}
\eea
where

\bea
E_1(t)&=&\Pi_i (1+\beta_it)^{(B_i-1)} \nn
F_1(t)&=&\int_0^t E_1(t')dt'
\l {eq:def3}
\eea
and

\be
B_i=\frac{r_i}{b_i}+1
\l {eq:def2}
\ee

\subsection{Fixed point structure}
\label{sec:3}
In the \sm if we keep only the dependence on the largest gauge
coupling, $g_3$, the equations have an infra-red-fixed point.
In this instance we have

\bea
E_1(t)&=&(1+\beta_3t)^{({B_3}-1)} \nn
F_1(t)&=&\frac{(1+\beta_3 t)^{B_3} }{{B_3} \beta_3} -\frac{1}{{B_3} \beta_3}
\l {eq:def1}
\eea

The solutions presented in the previous section do not make explicit
the fixed point structure. To do this it is necessary to eliminate the
functions $E_1$ and $F_1$ in eq(\r {eq:rgs})
giving

\be
\frac{Y_t(t)}{\tilde\alpha_3(t)}=\frac{Y_t(0)}{\tilde\alpha_3(0)}\; \; \;
 \frac{\R3^{B_3}}{1+\frac{sY_t(0)}
{{B_3}\beta_3}(\R3^{B_3}-1)}
\l {eq:intermediate}
\ee
After some algebra this reduces to
\be
\frac{Y_t(t)}{\tilde\alpha_3(t)}
=\left(\frac{Y_t}{\tilde\alpha_3}\right)^* \; \; \;
\frac{1}
{1+\RI3^{B_3}(\frac{Y_t}{\tilde\alpha_3})^*(\frac{\tilde\alpha_3(0)}{Y_t(0)}
-(\frac{\tilde\alpha_3}{Y_t})^*)}
\l {eq:soln1}
\ee
where the star superscript denotes the fixed point value
\be
\left(\frac{Y_t}{\tilde\alpha_3}\right)^*=\frac{{B_3} b_3}{s}
\ee
In this form it is clear that as $t\rightarrow\infty$ the ratio of the
Yukawa coupling to the gauge coupling tends to its fixed point value
because  $\RI3^{B_3}\rightarrow 0$. In the case of
the \sm
\bea
b_3&=&-7\nn
r&=&8 \nn
s&=&\frac{9}{2}
\eea
giving $(\frac{Y_t}{\tilde \alpha_3})^*_{SM}=\frac{2}{9}$
and $B_3=-\frac{1}{7}$. The smallness of the power $B_3$
in eq(\ref{eq:soln1}) together with the
slow evolution of the QCD coupling in the range $M_X$ to $m_t$ means
that in practice the fixed point is not reached.  If we take
$M_X=O(10^{16}GeV)$ then $\RI3^{B_3}\approx 0.8$ so we are quite far
away from the true fixed point. This means that, in general, one will
be sensitive to the initial value of the top Yukawa coupling.
For example if $\frac{Y_t(0)}{\tilde\alpha_3(0)}=
(\frac{Y_t}{\tilde\alpha_3})^*$ then it will remain at its fixed
point. On the other hand if
$\frac{Y_t(0)}{\tilde\alpha_3(0)}>>(\frac{Y_t}{\tilde\alpha_3})^*$
then we may see
from
eq(\r {eq:soln1}) that the value at low scales are relatively
insensitive to the initial value giving a ``Quasi-fixed-point'' value
which is really just the fixed point value radiatively corrected by known gauge
boson
contributions
\be
\left(\frac{Y_t}{\tilde\alpha_3}\right)^{QFP}=\frac{(\frac{Y_t}
{\tilde\alpha_3})^*}{(1-\RI3^{B_3})}
\l {eq:qfp}
\ee
Again using $M_X=O(10^{16}GeV)$ we find the Quasi-fixed-point gives a
value for the top quark Yukawa coupling and hence the top quark mass
approximately twice the true fixed point value, 220GeV rather than
110Gev\footnote{All masses quoted here are pole masses, related to
running masses by a simple correction \cite{pm}.}. In addition to the
QCD gauge
corrections just discussed which correct
for the fact that the RG flow is over a
relatively
small distance there are also significant corrections due to the
$SU(2)\otimes U(1)$ gauge interactions. In this case the
renormalisation group equations eq(\r {eq:rg}) do not have an exact
infra-red-fixed point and we must use the full solution to the
renormalisation group equations eq(\r {eq:rgs}). Including these
effects the quasi fixed point for the top pole mass is $240GeV$.

In the case of the minimal supersymmetric standard model (MSSM)
we have \cite{il,sir}
\bea
b_3&=&-3\nn
r&=&\frac{16}{3} \nn
s&=&6
\eea
resulting in a larger value for $(\frac{Y_t}{\tilde
\alpha_3})^*_{MSSM}
=\frac{7}{18}$ and ${B_3}$, ${B_3}=-\frac{7}{9}$. In this
case $\RI3^{B_3}\approx 0.46$ giving a somewhat closer approach to the
true fixed point; the Quasi-fixed-point is approximately
a third greater than the true fixed point at $155 sin \beta$
where $tan\beta$ is the ratio of the two
Higgs vacuum expectation values of the
MSSM. The effects of including $SU(2)\otimes U(1)$ corrections
is illustrated for the case of the MSSM by the graph of Fig \ref{fig:1}.
In this graph we plot the value of ($m_t/sin \beta$) versus the ratio
of the gauge to Yukawa couplings evaluated at the unification scale
normalised by their fixed point value. The effect of the quasi fixed
point is clear from the graph because the
region of large $Y_t(0)$ gives values of $m_t$ focused in a small
region (for $\frac{\tilde\alpha_3(0)}{Y_t(0)}<0.1
(\frac{\tilde\alpha_3}{Y_t})^*_{MSSM}$ we find
$205<\frac{m_t}{sin\beta}<210GeV$). Of course the interesting question
is whether the top mass is
determined by this (quasi) fixed point behaviour.
Until the value of $\sin\beta$ is determined this is not
known but clearly the large value predicted by the quasi-fixed-point
is needed to accommodate a top
mass of $O(174GeV)$.
\begin{figure}
\setlength{\unitlength}{0.240900pt}
\ifx\plotpoint\undefined\newsavebox{\plotpoint}\fi
\begin{picture}(1500,900)(-200,0)
\font\gnuplot=cmr10 at 10pt
\gnuplot
\sbox{\plotpoint}{\rule[-0.200pt]{0.400pt}{0.400pt}}%
\put(220.0,113.0){\rule[-0.200pt]{292.934pt}{0.400pt}}
\put(220.0,113.0){\rule[-0.200pt]{4.818pt}{0.400pt}}
\put(198,113){\makebox(0,0)[r]{0}}
\put(1416.0,113.0){\rule[-0.200pt]{4.818pt}{0.400pt}}
\put(220.0,215.0){\rule[-0.200pt]{4.818pt}{0.400pt}}
\put(198,215){\makebox(0,0)[r]{0.2}}
\put(1416.0,215.0){\rule[-0.200pt]{4.818pt}{0.400pt}}
\put(220.0,317.0){\rule[-0.200pt]{4.818pt}{0.400pt}}
\put(198,317){\makebox(0,0)[r]{0.4}}
\put(1416.0,317.0){\rule[-0.200pt]{4.818pt}{0.400pt}}
\put(220.0,419.0){\rule[-0.200pt]{4.818pt}{0.400pt}}
\put(198,419){\makebox(0,0)[r]{0.6}}
\put(1416.0,419.0){\rule[-0.200pt]{4.818pt}{0.400pt}}
\put(220.0,520.0){\rule[-0.200pt]{4.818pt}{0.400pt}}
\put(198,520){\makebox(0,0)[r]{0.8}}
\put(1416.0,520.0){\rule[-0.200pt]{4.818pt}{0.400pt}}
\put(220.0,622.0){\rule[-0.200pt]{4.818pt}{0.400pt}}
\put(198,622){\makebox(0,0)[r]{1}}
\put(1416.0,622.0){\rule[-0.200pt]{4.818pt}{0.400pt}}
\put(220.0,724.0){\rule[-0.200pt]{4.818pt}{0.400pt}}
\put(198,724){\makebox(0,0)[r]{1.2}}
\put(1416.0,724.0){\rule[-0.200pt]{4.818pt}{0.400pt}}
\put(220.0,826.0){\rule[-0.200pt]{4.818pt}{0.400pt}}
\put(198,826){\makebox(0,0)[r]{1.4}}
\put(1416.0,826.0){\rule[-0.200pt]{4.818pt}{0.400pt}}
\put(220.0,113.0){\rule[-0.200pt]{0.400pt}{4.818pt}}
\put(220,68){\makebox(0,0){160}}
\put(220.0,857.0){\rule[-0.200pt]{0.400pt}{4.818pt}}
\put(454.0,113.0){\rule[-0.200pt]{0.400pt}{4.818pt}}
\put(454,68){\makebox(0,0){170}}
\put(454.0,857.0){\rule[-0.200pt]{0.400pt}{4.818pt}}
\put(688.0,113.0){\rule[-0.200pt]{0.400pt}{4.818pt}}
\put(688,68){\makebox(0,0){180}}
\put(688.0,857.0){\rule[-0.200pt]{0.400pt}{4.818pt}}
\put(922.0,113.0){\rule[-0.200pt]{0.400pt}{4.818pt}}
\put(922,68){\makebox(0,0){190}}
\put(922.0,857.0){\rule[-0.200pt]{0.400pt}{4.818pt}}
\put(1155.0,113.0){\rule[-0.200pt]{0.400pt}{4.818pt}}
\put(1155,68){\makebox(0,0){200}}
\put(1155.0,857.0){\rule[-0.200pt]{0.400pt}{4.818pt}}
\put(1389.0,113.0){\rule[-0.200pt]{0.400pt}{4.818pt}}
\put(1389,68){\makebox(0,0){210}}
\put(1389.0,857.0){\rule[-0.200pt]{0.400pt}{4.818pt}}
\put(220.0,113.0){\rule[-0.200pt]{292.934pt}{0.400pt}}
\put(1436.0,113.0){\rule[-0.200pt]{0.400pt}{184.048pt}}
\put(220.0,877.0){\rule[-0.200pt]{292.934pt}{0.400pt}}
\put(-65,495){\makebox(0,0){$\left(\frac{\tilde\alpha(0)}{Y_t(0)}
\right)/ \left(\frac{\tilde\alpha}{Y_t}\right)_{MSSM}^*$}}
\put(828,23){\makebox(0,0){$m_t / sin \beta$ (GeV)}}
\put(220.0,113.0){\rule[-0.200pt]{0.400pt}{184.048pt}}
\put(220,809){\usebox{\plotpoint}}
\multiput(220.00,807.92)(0.543,-0.492){19}{\rule{0.536pt}{0.118pt}}
\multiput(220.00,808.17)(10.887,-11.000){2}{\rule{0.268pt}{0.400pt}}
\multiput(232.00,796.92)(0.590,-0.492){19}{\rule{0.573pt}{0.118pt}}
\multiput(232.00,797.17)(11.811,-11.000){2}{\rule{0.286pt}{0.400pt}}
\multiput(245.00,785.92)(0.543,-0.492){19}{\rule{0.536pt}{0.118pt}}
\multiput(245.00,786.17)(10.887,-11.000){2}{\rule{0.268pt}{0.400pt}}
\multiput(257.00,774.92)(0.600,-0.491){17}{\rule{0.580pt}{0.118pt}}
\multiput(257.00,775.17)(10.796,-10.000){2}{\rule{0.290pt}{0.400pt}}
\multiput(269.00,764.92)(0.543,-0.492){19}{\rule{0.536pt}{0.118pt}}
\multiput(269.00,765.17)(10.887,-11.000){2}{\rule{0.268pt}{0.400pt}}
\multiput(281.00,753.92)(0.652,-0.491){17}{\rule{0.620pt}{0.118pt}}
\multiput(281.00,754.17)(11.713,-10.000){2}{\rule{0.310pt}{0.400pt}}
\multiput(294.00,743.92)(0.600,-0.491){17}{\rule{0.580pt}{0.118pt}}
\multiput(294.00,744.17)(10.796,-10.000){2}{\rule{0.290pt}{0.400pt}}
\multiput(306.00,733.92)(0.600,-0.491){17}{\rule{0.580pt}{0.118pt}}
\multiput(306.00,734.17)(10.796,-10.000){2}{\rule{0.290pt}{0.400pt}}
\multiput(318.00,723.92)(0.652,-0.491){17}{\rule{0.620pt}{0.118pt}}
\multiput(318.00,724.17)(11.713,-10.000){2}{\rule{0.310pt}{0.400pt}}
\multiput(331.00,713.92)(0.600,-0.491){17}{\rule{0.580pt}{0.118pt}}
\multiput(331.00,714.17)(10.796,-10.000){2}{\rule{0.290pt}{0.400pt}}
\multiput(343.00,703.92)(0.600,-0.491){17}{\rule{0.580pt}{0.118pt}}
\multiput(343.00,704.17)(10.796,-10.000){2}{\rule{0.290pt}{0.400pt}}
\multiput(355.00,693.92)(0.600,-0.491){17}{\rule{0.580pt}{0.118pt}}
\multiput(355.00,694.17)(10.796,-10.000){2}{\rule{0.290pt}{0.400pt}}
\multiput(367.00,683.92)(0.652,-0.491){17}{\rule{0.620pt}{0.118pt}}
\multiput(367.00,684.17)(11.713,-10.000){2}{\rule{0.310pt}{0.400pt}}
\multiput(380.00,673.93)(0.669,-0.489){15}{\rule{0.633pt}{0.118pt}}
\multiput(380.00,674.17)(10.685,-9.000){2}{\rule{0.317pt}{0.400pt}}
\multiput(392.00,664.92)(0.600,-0.491){17}{\rule{0.580pt}{0.118pt}}
\multiput(392.00,665.17)(10.796,-10.000){2}{\rule{0.290pt}{0.400pt}}
\multiput(404.00,654.93)(0.728,-0.489){15}{\rule{0.678pt}{0.118pt}}
\multiput(404.00,655.17)(11.593,-9.000){2}{\rule{0.339pt}{0.400pt}}
\multiput(417.00,645.92)(0.600,-0.491){17}{\rule{0.580pt}{0.118pt}}
\multiput(417.00,646.17)(10.796,-10.000){2}{\rule{0.290pt}{0.400pt}}
\multiput(429.00,635.93)(0.669,-0.489){15}{\rule{0.633pt}{0.118pt}}
\multiput(429.00,636.17)(10.685,-9.000){2}{\rule{0.317pt}{0.400pt}}
\multiput(441.00,626.93)(0.669,-0.489){15}{\rule{0.633pt}{0.118pt}}
\multiput(441.00,627.17)(10.685,-9.000){2}{\rule{0.317pt}{0.400pt}}
\multiput(453.00,617.93)(0.728,-0.489){15}{\rule{0.678pt}{0.118pt}}
\multiput(453.00,618.17)(11.593,-9.000){2}{\rule{0.339pt}{0.400pt}}
\multiput(466.00,608.93)(0.669,-0.489){15}{\rule{0.633pt}{0.118pt}}
\multiput(466.00,609.17)(10.685,-9.000){2}{\rule{0.317pt}{0.400pt}}
\multiput(478.00,599.93)(0.669,-0.489){15}{\rule{0.633pt}{0.118pt}}
\multiput(478.00,600.17)(10.685,-9.000){2}{\rule{0.317pt}{0.400pt}}
\multiput(490.00,590.93)(0.728,-0.489){15}{\rule{0.678pt}{0.118pt}}
\multiput(490.00,591.17)(11.593,-9.000){2}{\rule{0.339pt}{0.400pt}}
\multiput(503.00,581.93)(0.669,-0.489){15}{\rule{0.633pt}{0.118pt}}
\multiput(503.00,582.17)(10.685,-9.000){2}{\rule{0.317pt}{0.400pt}}
\multiput(515.00,572.93)(0.758,-0.488){13}{\rule{0.700pt}{0.117pt}}
\multiput(515.00,573.17)(10.547,-8.000){2}{\rule{0.350pt}{0.400pt}}
\multiput(527.00,564.93)(0.669,-0.489){15}{\rule{0.633pt}{0.118pt}}
\multiput(527.00,565.17)(10.685,-9.000){2}{\rule{0.317pt}{0.400pt}}
\multiput(539.00,555.93)(0.824,-0.488){13}{\rule{0.750pt}{0.117pt}}
\multiput(539.00,556.17)(11.443,-8.000){2}{\rule{0.375pt}{0.400pt}}
\multiput(552.00,547.93)(0.669,-0.489){15}{\rule{0.633pt}{0.118pt}}
\multiput(552.00,548.17)(10.685,-9.000){2}{\rule{0.317pt}{0.400pt}}
\multiput(564.00,538.93)(0.758,-0.488){13}{\rule{0.700pt}{0.117pt}}
\multiput(564.00,539.17)(10.547,-8.000){2}{\rule{0.350pt}{0.400pt}}
\multiput(576.00,530.93)(0.758,-0.488){13}{\rule{0.700pt}{0.117pt}}
\multiput(576.00,531.17)(10.547,-8.000){2}{\rule{0.350pt}{0.400pt}}
\multiput(588.00,522.93)(0.728,-0.489){15}{\rule{0.678pt}{0.118pt}}
\multiput(588.00,523.17)(11.593,-9.000){2}{\rule{0.339pt}{0.400pt}}
\multiput(601.00,513.93)(0.758,-0.488){13}{\rule{0.700pt}{0.117pt}}
\multiput(601.00,514.17)(10.547,-8.000){2}{\rule{0.350pt}{0.400pt}}
\multiput(613.00,505.93)(0.758,-0.488){13}{\rule{0.700pt}{0.117pt}}
\multiput(613.00,506.17)(10.547,-8.000){2}{\rule{0.350pt}{0.400pt}}
\multiput(625.00,497.93)(0.824,-0.488){13}{\rule{0.750pt}{0.117pt}}
\multiput(625.00,498.17)(11.443,-8.000){2}{\rule{0.375pt}{0.400pt}}
\multiput(638.00,489.93)(0.758,-0.488){13}{\rule{0.700pt}{0.117pt}}
\multiput(638.00,490.17)(10.547,-8.000){2}{\rule{0.350pt}{0.400pt}}
\multiput(650.00,481.93)(0.874,-0.485){11}{\rule{0.786pt}{0.117pt}}
\multiput(650.00,482.17)(10.369,-7.000){2}{\rule{0.393pt}{0.400pt}}
\multiput(662.00,474.93)(0.758,-0.488){13}{\rule{0.700pt}{0.117pt}}
\multiput(662.00,475.17)(10.547,-8.000){2}{\rule{0.350pt}{0.400pt}}
\multiput(674.00,466.93)(0.824,-0.488){13}{\rule{0.750pt}{0.117pt}}
\multiput(674.00,467.17)(11.443,-8.000){2}{\rule{0.375pt}{0.400pt}}
\multiput(687.00,458.93)(0.758,-0.488){13}{\rule{0.700pt}{0.117pt}}
\multiput(687.00,459.17)(10.547,-8.000){2}{\rule{0.350pt}{0.400pt}}
\multiput(699.00,450.93)(0.874,-0.485){11}{\rule{0.786pt}{0.117pt}}
\multiput(699.00,451.17)(10.369,-7.000){2}{\rule{0.393pt}{0.400pt}}
\multiput(711.00,443.93)(0.824,-0.488){13}{\rule{0.750pt}{0.117pt}}
\multiput(711.00,444.17)(11.443,-8.000){2}{\rule{0.375pt}{0.400pt}}
\multiput(724.00,435.93)(0.874,-0.485){11}{\rule{0.786pt}{0.117pt}}
\multiput(724.00,436.17)(10.369,-7.000){2}{\rule{0.393pt}{0.400pt}}
\multiput(736.00,428.93)(0.758,-0.488){13}{\rule{0.700pt}{0.117pt}}
\multiput(736.00,429.17)(10.547,-8.000){2}{\rule{0.350pt}{0.400pt}}
\multiput(748.00,420.93)(0.874,-0.485){11}{\rule{0.786pt}{0.117pt}}
\multiput(748.00,421.17)(10.369,-7.000){2}{\rule{0.393pt}{0.400pt}}
\multiput(760.00,413.93)(0.950,-0.485){11}{\rule{0.843pt}{0.117pt}}
\multiput(760.00,414.17)(11.251,-7.000){2}{\rule{0.421pt}{0.400pt}}
\multiput(773.00,406.93)(0.874,-0.485){11}{\rule{0.786pt}{0.117pt}}
\multiput(773.00,407.17)(10.369,-7.000){2}{\rule{0.393pt}{0.400pt}}
\multiput(785.00,399.93)(0.874,-0.485){11}{\rule{0.786pt}{0.117pt}}
\multiput(785.00,400.17)(10.369,-7.000){2}{\rule{0.393pt}{0.400pt}}
\multiput(797.00,392.93)(0.824,-0.488){13}{\rule{0.750pt}{0.117pt}}
\multiput(797.00,393.17)(11.443,-8.000){2}{\rule{0.375pt}{0.400pt}}
\multiput(810.00,384.93)(0.874,-0.485){11}{\rule{0.786pt}{0.117pt}}
\multiput(810.00,385.17)(10.369,-7.000){2}{\rule{0.393pt}{0.400pt}}
\multiput(822.00,377.93)(1.033,-0.482){9}{\rule{0.900pt}{0.116pt}}
\multiput(822.00,378.17)(10.132,-6.000){2}{\rule{0.450pt}{0.400pt}}
\multiput(834.00,371.93)(0.874,-0.485){11}{\rule{0.786pt}{0.117pt}}
\multiput(834.00,372.17)(10.369,-7.000){2}{\rule{0.393pt}{0.400pt}}
\multiput(846.00,364.93)(0.950,-0.485){11}{\rule{0.843pt}{0.117pt}}
\multiput(846.00,365.17)(11.251,-7.000){2}{\rule{0.421pt}{0.400pt}}
\multiput(859.00,357.93)(0.874,-0.485){11}{\rule{0.786pt}{0.117pt}}
\multiput(859.00,358.17)(10.369,-7.000){2}{\rule{0.393pt}{0.400pt}}
\multiput(871.00,350.93)(0.874,-0.485){11}{\rule{0.786pt}{0.117pt}}
\multiput(871.00,351.17)(10.369,-7.000){2}{\rule{0.393pt}{0.400pt}}
\multiput(883.00,343.93)(1.123,-0.482){9}{\rule{0.967pt}{0.116pt}}
\multiput(883.00,344.17)(10.994,-6.000){2}{\rule{0.483pt}{0.400pt}}
\multiput(896.00,337.93)(0.874,-0.485){11}{\rule{0.786pt}{0.117pt}}
\multiput(896.00,338.17)(10.369,-7.000){2}{\rule{0.393pt}{0.400pt}}
\multiput(908.00,330.93)(0.874,-0.485){11}{\rule{0.786pt}{0.117pt}}
\multiput(908.00,331.17)(10.369,-7.000){2}{\rule{0.393pt}{0.400pt}}
\multiput(920.00,323.93)(1.033,-0.482){9}{\rule{0.900pt}{0.116pt}}
\multiput(920.00,324.17)(10.132,-6.000){2}{\rule{0.450pt}{0.400pt}}
\multiput(932.00,317.93)(0.950,-0.485){11}{\rule{0.843pt}{0.117pt}}
\multiput(932.00,318.17)(11.251,-7.000){2}{\rule{0.421pt}{0.400pt}}
\multiput(945.00,310.93)(1.033,-0.482){9}{\rule{0.900pt}{0.116pt}}
\multiput(945.00,311.17)(10.132,-6.000){2}{\rule{0.450pt}{0.400pt}}
\multiput(957.00,304.93)(1.033,-0.482){9}{\rule{0.900pt}{0.116pt}}
\multiput(957.00,305.17)(10.132,-6.000){2}{\rule{0.450pt}{0.400pt}}
\multiput(969.00,298.93)(0.950,-0.485){11}{\rule{0.843pt}{0.117pt}}
\multiput(969.00,299.17)(11.251,-7.000){2}{\rule{0.421pt}{0.400pt}}
\multiput(982.00,291.93)(1.033,-0.482){9}{\rule{0.900pt}{0.116pt}}
\multiput(982.00,292.17)(10.132,-6.000){2}{\rule{0.450pt}{0.400pt}}
\multiput(994.00,285.93)(1.033,-0.482){9}{\rule{0.900pt}{0.116pt}}
\multiput(994.00,286.17)(10.132,-6.000){2}{\rule{0.450pt}{0.400pt}}
\multiput(1006.00,279.93)(1.033,-0.482){9}{\rule{0.900pt}{0.116pt}}
\multiput(1006.00,280.17)(10.132,-6.000){2}{\rule{0.450pt}{0.400pt}}
\multiput(1018.00,273.93)(1.123,-0.482){9}{\rule{0.967pt}{0.116pt}}
\multiput(1018.00,274.17)(10.994,-6.000){2}{\rule{0.483pt}{0.400pt}}
\multiput(1031.00,267.93)(1.033,-0.482){9}{\rule{0.900pt}{0.116pt}}
\multiput(1031.00,268.17)(10.132,-6.000){2}{\rule{0.450pt}{0.400pt}}
\multiput(1043.00,261.93)(1.033,-0.482){9}{\rule{0.900pt}{0.116pt}}
\multiput(1043.00,262.17)(10.132,-6.000){2}{\rule{0.450pt}{0.400pt}}
\multiput(1055.00,255.93)(1.123,-0.482){9}{\rule{0.967pt}{0.116pt}}
\multiput(1055.00,256.17)(10.994,-6.000){2}{\rule{0.483pt}{0.400pt}}
\multiput(1068.00,249.93)(1.033,-0.482){9}{\rule{0.900pt}{0.116pt}}
\multiput(1068.00,250.17)(10.132,-6.000){2}{\rule{0.450pt}{0.400pt}}
\multiput(1080.00,243.93)(1.033,-0.482){9}{\rule{0.900pt}{0.116pt}}
\multiput(1080.00,244.17)(10.132,-6.000){2}{\rule{0.450pt}{0.400pt}}
\multiput(1092.00,237.93)(1.033,-0.482){9}{\rule{0.900pt}{0.116pt}}
\multiput(1092.00,238.17)(10.132,-6.000){2}{\rule{0.450pt}{0.400pt}}
\multiput(1104.00,231.93)(1.378,-0.477){7}{\rule{1.140pt}{0.115pt}}
\multiput(1104.00,232.17)(10.634,-5.000){2}{\rule{0.570pt}{0.400pt}}
\multiput(1117.00,226.93)(1.033,-0.482){9}{\rule{0.900pt}{0.116pt}}
\multiput(1117.00,227.17)(10.132,-6.000){2}{\rule{0.450pt}{0.400pt}}
\multiput(1129.00,220.93)(1.033,-0.482){9}{\rule{0.900pt}{0.116pt}}
\multiput(1129.00,221.17)(10.132,-6.000){2}{\rule{0.450pt}{0.400pt}}
\multiput(1141.00,214.93)(1.267,-0.477){7}{\rule{1.060pt}{0.115pt}}
\multiput(1141.00,215.17)(9.800,-5.000){2}{\rule{0.530pt}{0.400pt}}
\multiput(1153.00,209.93)(1.123,-0.482){9}{\rule{0.967pt}{0.116pt}}
\multiput(1153.00,210.17)(10.994,-6.000){2}{\rule{0.483pt}{0.400pt}}
\multiput(1166.00,203.93)(1.033,-0.482){9}{\rule{0.900pt}{0.116pt}}
\multiput(1166.00,204.17)(10.132,-6.000){2}{\rule{0.450pt}{0.400pt}}
\multiput(1178.00,197.93)(1.267,-0.477){7}{\rule{1.060pt}{0.115pt}}
\multiput(1178.00,198.17)(9.800,-5.000){2}{\rule{0.530pt}{0.400pt}}
\multiput(1190.00,192.93)(1.123,-0.482){9}{\rule{0.967pt}{0.116pt}}
\multiput(1190.00,193.17)(10.994,-6.000){2}{\rule{0.483pt}{0.400pt}}
\multiput(1203.00,186.93)(1.267,-0.477){7}{\rule{1.060pt}{0.115pt}}
\multiput(1203.00,187.17)(9.800,-5.000){2}{\rule{0.530pt}{0.400pt}}
\multiput(1215.00,181.93)(1.267,-0.477){7}{\rule{1.060pt}{0.115pt}}
\multiput(1215.00,182.17)(9.800,-5.000){2}{\rule{0.530pt}{0.400pt}}
\multiput(1227.00,176.93)(1.033,-0.482){9}{\rule{0.900pt}{0.116pt}}
\multiput(1227.00,177.17)(10.132,-6.000){2}{\rule{0.450pt}{0.400pt}}
\multiput(1239.00,170.93)(1.378,-0.477){7}{\rule{1.140pt}{0.115pt}}
\multiput(1239.00,171.17)(10.634,-5.000){2}{\rule{0.570pt}{0.400pt}}
\multiput(1252.00,165.93)(1.267,-0.477){7}{\rule{1.060pt}{0.115pt}}
\multiput(1252.00,166.17)(9.800,-5.000){2}{\rule{0.530pt}{0.400pt}}
\multiput(1264.00,160.93)(1.267,-0.477){7}{\rule{1.060pt}{0.115pt}}
\multiput(1264.00,161.17)(9.800,-5.000){2}{\rule{0.530pt}{0.400pt}}
\multiput(1276.00,155.93)(1.123,-0.482){9}{\rule{0.967pt}{0.116pt}}
\multiput(1276.00,156.17)(10.994,-6.000){2}{\rule{0.483pt}{0.400pt}}
\multiput(1289.00,149.93)(1.267,-0.477){7}{\rule{1.060pt}{0.115pt}}
\multiput(1289.00,150.17)(9.800,-5.000){2}{\rule{0.530pt}{0.400pt}}
\multiput(1301.00,144.93)(1.267,-0.477){7}{\rule{1.060pt}{0.115pt}}
\multiput(1301.00,145.17)(9.800,-5.000){2}{\rule{0.530pt}{0.400pt}}
\multiput(1313.00,139.93)(1.267,-0.477){7}{\rule{1.060pt}{0.115pt}}
\multiput(1313.00,140.17)(9.800,-5.000){2}{\rule{0.530pt}{0.400pt}}
\multiput(1325.00,134.93)(1.378,-0.477){7}{\rule{1.140pt}{0.115pt}}
\multiput(1325.00,135.17)(10.634,-5.000){2}{\rule{0.570pt}{0.400pt}}
\multiput(1338.00,129.93)(1.267,-0.477){7}{\rule{1.060pt}{0.115pt}}
\multiput(1338.00,130.17)(9.800,-5.000){2}{\rule{0.530pt}{0.400pt}}
\multiput(1350.00,124.93)(1.267,-0.477){7}{\rule{1.060pt}{0.115pt}}
\multiput(1350.00,125.17)(9.800,-5.000){2}{\rule{0.530pt}{0.400pt}}
\multiput(1362.00,119.93)(1.378,-0.477){7}{\rule{1.140pt}{0.115pt}}
\multiput(1362.00,120.17)(10.634,-5.000){2}{\rule{0.570pt}{0.400pt}}
\multiput(1375.00,114.95)(1.579,-0.447){3}{\rule{1.167pt}{0.108pt}}
\multiput(1375.00,115.17)(5.579,-3.000){2}{\rule{0.583pt}{0.400pt}}
\l{fig:1}
\end{picture}
\caption{ Plot showing the running top quark pole mass as a function
of the initial value of the Yukawa coupling.}
\end{figure}
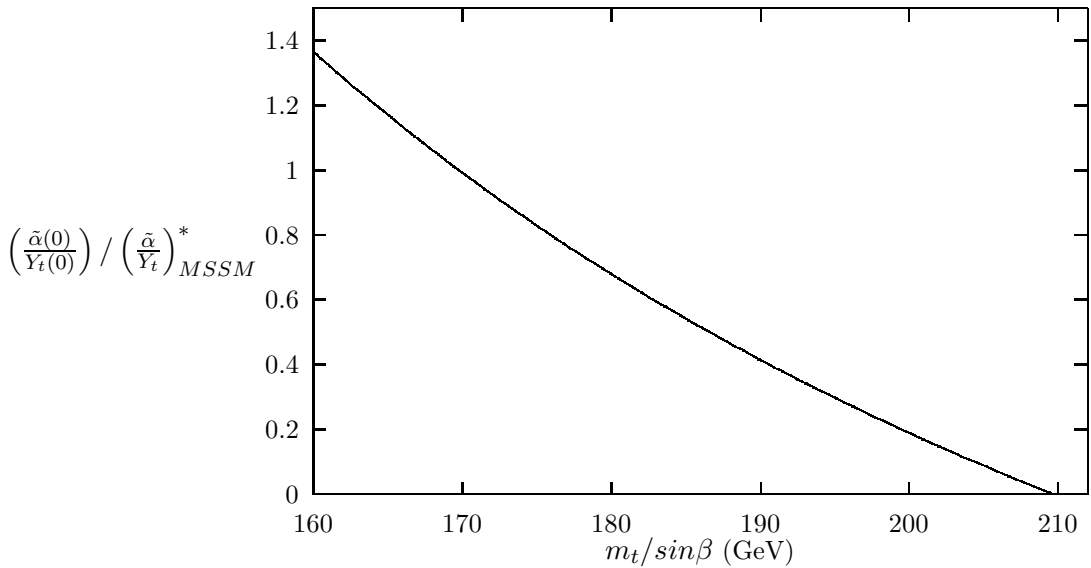

\section{Are ``Unified'' fixed points relevant to the determination of
quark and lepton masses
and mixing angles?}
\label {sec:comp}
As we have seen the quasi-infra-red fixed point may be of importance
in determining the top quark mass in supersymmetric theories. The
difficulty in being more definite lies partly with the uncertainty in
$\beta$ and partly in the fact that (cf. Fig \ref{fig:1} ) the value
of the Yukawa coupling at low energies is not completely insensitive
to the initial value. The former difficulty will be eliminated when
(and if) the structure of the low-energy effective supersymmetric
theory is determined experimentally and $tan \beta $ measured.
We shall consider here whether
the {\it fixed point} structure of the underlying theory beyond
the \sm sheds light on the
initial value at the gauge unification scale and hence the latter
difficulty. The obvious attraction of this would be that the couplings
would be determined simply in terms of the gauge couplings by the
dynamics and knowlege of the gauge group and multiplet content would
suffice to determine the
couplings.

At first sight it appears that the infra-red structure of the theory
beyond the \sm will play no significant r$\hat{o}$le in the
determination of the parameters at the gauge unification scale because
the domain over which the renormalisation group flow is relevant is
too small to give appreciable approach to any infra-red fixed points.
This may be seen clearly in the solution of Section \r{sec:3}. From
eq(\r {eq:soln1}) we see that the important parameter determining
whether the fixed point is reached or not is \EP. Given that the maximum
range of t over which the renormalisation flow can be computed runs
only from the compactification scale, $\Lambda_c$,
(or the Planck scale, $M_P$, which is expected to be slightly
larger\footnote{Above the Planck scale unknown gravitational effects
are important.}) to the
gauge unification scale, $M_X\approx 10^{16}GeV$, we see that  $t\le
ln((\Lambda_c/M_X)^2)\approx 9$, much less than the value $t\approx
65$ that is relevant in the flow from the gauge unification scale to
the electroweak
breaking scale! As a result we might expect that
$(\frac{\alpha(M_X)}{\alpha(M_C)})^B$ is not small
and that the infra-red-fixed point plays a negligible
role in determining the couplings at the gauge unification scale.

However we shall argue that there are  reasons why this conclusion
 is likely be wrong in many extensions of the \smp Firstly the value of
$(\frac{\alpha(M_X)}{\alpha(M_C)})^B$
depends on the exponent B which,
as we shall demonstrate by specific examples, is likely to be much
larger than in the \sm or MSSM. Secondly the ratio of couplings under the
exponent depends not only on $t$ but also on the beta function above
gauge unification. Again, as we shall demonstrate by considering
realistic examples, this is expected to be larger than that of the
MSSM. These effects act in the same direction and can lead to the fixed points
structure of the theory beyond the \sm being {\it more} important than
in the \sm or
the MSSM.
There is an even more compelling reason for the importance of the
fixed point structure which applies to theories with a stage of
compactification. In this case the evolution of couplings is much
faster
(following a power law rather than a logarithmic evolution) leading
to a very small value for
$(\frac{\alpha(M_C)}{\alpha(M_P)})^B$.
We shall illustrate these points by several representative(?) examples.

We first consider some phenomenologically interesting
extensions of the \sm which have infra-red-fixed points for the ratio
of gauge to Yukawa couplings. These examples apply in uncompactified
theories or, if there is a stage of compactification, below the
compactification
scale. Following from the discussion of Section
\r {sec:3}, realisation of this idea requires a theory with a single
gauge group or a product of identical gauge groups coupling to the
quark fields for otherwise the gauge couplings will (in all
probability) have different $\beta$ functions and hence will not have
infra-red-fixed points for the ratio of gauge to Yukawa
couplings\footnote{Also, more seriously, theories with more than one
gauge coupling will spoil the success of the minimal unification of
gauge couplings due to the
additional running of gauge couplings above $M_X$.}. We
will consider several examples of such a theory namely supersymmetric
\footnote{We restrict our attention to supersymmetric theories to
preserve the success of the unification predictions for gauge
couplings and quark and lepton masses but the general structure
considered here applies equally well to non-supersymmetric
theories.} $SU(5)$, $SO(10)$ and $SU(3)\otimes SU(3) \otimes SU(3)$.
The latter is the simplest such extension of the \sm and we consider it first.

\subsection{$SU(3)^3$}
The group $SU(3)^3$ has many attractions as a non-Grand-Unified
extension of the \smp Provided the multiplet content is chosen
symmetrically, the gauge couplings above the $SU(3)^3$ unification
scale, $M_X$, evolve together. As a result it offers an example of a
theory which can be embedded in the superstring which preserves the
success of the unification predictions for gauge couplings even if, as
has been found to be usuually the case in the compactified string
theories so far analysed, the unification or compactification
scale is much higher than $10^{16}GeV$. It also has a relatively
simple breaking pattern taking it to the \sm using just the
fundamental representations of Higgs fields and this means it fits
nicely into superstrings built using level-1 Kac-Moody algebras.
Indeed there are 3-generation examples known of semi-realistic
compactified string theories with the
gauge group $SU(3)^3$.

For our purposes here it is not necessary to know the string origin
(if any) for, as we stressed above, the fixed point structure
determines the couplings of the theory. All that is needed is the
multiplet content and the remaining (discrete) symmetry structure
which dictates the allowed couplings.  The light multiplet structure after
symmetry breaking is just that of the MSSM.  Shafi
et. al. \cite{shafi} have shown how this can naturally come from a
model with $SU(3)^3$ symmetry
through discrete symmetries.  The multiplet content before symmetry
breaking we take to
consist of  $n_g$ families contained in $n_g$ copies of $I$
representations where $I=((1,3,\bar{3})+(\bar{3},1,3)+(3,\bar{3},1))$.
In addition there are two further copies of $I$ representations which
contain the Higgs fields. The renormalisation group equations
for this theory are given by
\bea
\frac{d \tilde \alpha_i}{d t} &=& -6 \tilde \alpha_i^2 \nn
\frac{d Y_t}{d t}&=& (16\tilde\alpha - 9 Y_t)Y_t
\eea
where in the second equation we have used the fact that all three
gauge couplings are equal,  $\tilde\alpha_i=\tilde\alpha$,
(assuming equal initial values).
Note that the gauge couplings are not asymptotically free due to the
profusion of matter fields. This we believe is a very common feature
of extensions of the \sm which include all the fields necessary to
break the symmetry fully. We will return to the implications of this
shortly.

Applying  the results derived above we see that there is indeed  a
fixed point given by
$(\frac{Y_t}{\tilde\alpha})^*_{SU(3)^3}=\frac{22}{9}$ and
$B=\frac{11}{3}$,
much larger than in the cases considered above.
The important factor determining the rate of approach to the fixed point is
$(\frac{\alpha(M_X)}{\alpha(M_C)})^B$ and because above $M_X$
the theory is not asymptotically free the larger B makes this factor smaller.
In addition the factor also decreases because the running coupling
evolves faster due to the larger $\beta$ function. If
we take the conservative view and evolve from the compactification
scale $O(10^{18}GeV)$ to the gauge unification scale $O(10^{16})GeV$,
normalising the gauge coupling to the MSSM gauge unification value
$\alpha^{-1} (M_X)\approx 24$, we find \EPP$\approx 0.48$.
 Thus, the fixed point in the evolution between $10^{18}GeV$ to
$10^{16}GeV$ plays a role as important in the $SU(3)^3$ theory as
does in the MSSM evolving
between $10^{16}GeV$ and $10^{2}GeV$!
Following the analysis of Section \ref{sec:3} we
expect the Yukawa coupling, for large initial values at
the compactification scale, to be some
30\% larger than the fixed point value giving, from eq(\ref{eq:qfp})
$\frac{Y_t(M_X)}{\tilde\alpha(M_X)}\approx 4.2$. In terms of the MSSM
fixed point value plotted in Fig \r{fig:1} this corresponds to
\FPRX$\approx 0.08$ which means the low energy value will be very close
to the quasi fixed point value.  The effect of this on the final
prediction for the top quark mass may be quantified by noting that
if the initial value of \FPRP is $x$ then

\bea
x'&=&\frac{\left( \frac{\tilde \alpha(M_X)}{Y_t(M_X)} \right) }
{\left( \frac{\tilde\alpha}{Y_t}\right)_{MSSM}^* }\nn
& = &x
\left( \frac{\tilde\alpha(M_X)}{\tilde\alpha(M_C)} \right)^B
+ \frac{\left(\frac{\tilde\alpha}
{Y_t}\right)_{SU(3)^3}^*\left( 1 - \left( \frac{\tilde\alpha(M_X)}
{\tilde\alpha(M_C)}\right)^B\right)}{
\left(\frac{\tilde\alpha}{Y_t}\right)_{MSSM}^*}
\l{eq:focus}
\eea
Small $x'$ means the Yukawa coupling is in the domain of attraction
of the quasi fixed point of the MSSM.
If the second term is the largest then we see we are already in this
domain since
$(\frac{\alpha}{Y})^*_{SU(3)^3}/(\frac{\alpha}{Y})^*_{MSSM})<<1$. If
the first term is largest there is still a focusing effect taking the
coupling
towards the domain of attraction because eq(\ref{eq:focus}) gives
approximately $x'=x (\frac{\alpha(M_X)}{\alpha(M_C)})^B$. To
illustrate the effect note that if $x \leq 1/2$ at $M_X$ the range of
values for the
top mass is $186 GeV \leq  m_t \leq 210 GeV $;
if $x \leq 1/2$ at $M_C$ after focusing the range of values becomes
$ 195 GeV \leq m_t \leq 210 GeV$.

Let us consider how stable is our result to changes in the
structure of the $SU(3)^3$ theory.  In many string compactifications
there arise additional states in conjugate representations which
acquire mass at the stage of gauge symmetry breaking and do not appear
in the low energy theory. Let us consider the effect of such states by
adding to our theory n copies of chiral superfields in $(I+\bar{I})$
representations\footnote{In the Tian Yau three generation
theory n=6 \cite{ty}.}. First we consider the case that their Yukawa
couplings are small and so the only effect of these fields is to
change the gauge beta function $b_i=6+6n$. This in turn affects the
position of the fixed point giving
$(\frac{Y_t}{\tilde\alpha})^*=\frac{22+6n}{9}$. Note the systematic
effect of adding additional matter is to {\it increase } $Y_t$ driving
it closer to the quasi fixed point. It also changes
(reduces) B while increasing the rate of change of the running
coupling, these changes going in opposite directions in the
determination of $(\frac{\alpha(M_X)}{\alpha(M_C)})^B$.
Putting this together gives
$(\frac{\alpha(M_X)}{\alpha(M_C)})^B = 0.22, 0.02$ for
$n=2,4$ respectively showing that the effect of additional matter
is to speed up the approach to the fixed point. The results are
summarised in Table \ref{table:1}. In particular note that for
the case of n=4 or more the ratio of couplings sare within 1\%
of the fixed point value.

\begin{table}
\begin{center}
\begin{tabular}{|c|c|c|c|}\hline
$n$  &  $(\frac{Y_t}{\tilde\alpha})^*$ & \EPP  &
$(\frac{\tilde\alpha}{Y_t})_{SU(3)^3}^*/(\frac{\tilde\alpha}{Y_t})_{MSSM}^*$
\\ \hline\hline
0    & 2.44 & 0.48 & 0.16 \\ \hline
2    & 3.78 & 0.22 & 0.10 \\ \hline
4    & 5.11 & 0.02 & 0.08 \\ \hline
\end{tabular}
\end{center}
\l{table:1}
\caption{$SU(3)^3$}
\end{table}

\subsection{$SU(5)$}
In our next example we will consider an SU(5) model. Matter
is arranged in three generations in $I$=
$\{ \psi^{xy}(10) + \phi_x(\overline{5})\}$ representations together
with n further copies of chiral superfields in ($I+\bar{I}$)
representations
plus a Higgs sector made
up of a (complex) adjoint, $\Sigma(24)$, to break SU(5) and a set of
Weinberg-Salam 5-plets $H_1(5) + H_2(\overline{5})$. Keeping only the
Yukawa coupling leading to the top quark mass the renormalization group
equations are given by:
\bea
\frac{d \tilde \alpha} {dt} &=& (3-4n) \tilde \alpha ^2 \nn
\frac{d Y_t}{dt} &=& (2(\frac{48}{5}) \tilde \alpha - 9Y_t)Y_t
\eea
The gauge coupling
is asymptotically free only for n=0. Then, the fixed points are given by:

\be
\left(\frac{Y_t}{\tilde\alpha}\right)^* = \frac{2(\frac{48}{5}) - 3 + 4n}{9}
\ee
For n=0,  $B=-\frac{27}{5}$
and evolving from the compactification scale to the gauge unification
scale we get \EPP $\approx 0.62$.
It can be seen that in this model, the fixed
point structure plays a slightly less important role than in the {$SU(3)^3$}
case. However for $n\ne 0$ the approach to the fixed point is more
rapid as is shown in Table(\ref{table:2}) and essentially reaches the
fixed point for $n\ge 8$. In all cases the fixed point value is in the
domain of
attraction of the quasi fixed point of the MSSM.

\begin{table}
\begin{center}
\begin{tabular}{|c|c|c|c|} \hline
$n$ & $(\frac{Y_t}{\tilde\alpha})^*$ & \EPP &
$(\frac{\tilde\alpha}{Y_t})_{SU(5)}^*/(\frac{\tilde\alpha}{Y_t})_{MSSM}^*$
\\ \hline\hline
0 & 1.80 & 0.62 & 0.22 \\ \hline
2 & 2.68 & 0.45 & 0.15 \\ \hline
4 & 3.57 & 0.29 & 0.11 \\ \hline
8 & 5.36 & 0.03 & 0.07 \\ \hline
\end{tabular}
\end{center}
\l{table:2}
\caption{SU(5)}
\end{table}

\subsection{$SO(10)$}

One of the advantages of SO(10)
over SU(5) Grand Unification is that only one 16-dimensional
spinor representation of SO(10) is needed to
accomodate all fermions (including the right handed neutrino)
of one generation. Unlike SU(5), SO(10) is a group of rank 5
with the extra diagonal generator of SO(10) being B - L as in
the left-right symmetric groups.
Because of this there exist
several intermediate symmetries through which SO(10) can descend to the
$SU(3) \otimes SU(2)\otimes U(1)$ group.
For this paper we consider the simplest route
which is given by the following
stages of symmetry breaking:

\be
\begin{array}{ccccccc}
SO(10) &\longrightarrow & SU(5) &\longrightarrow &
SU(3)\otimes SU(2)\otimes U(1) &
\longrightarrow & SU(3)\otimes U(1) \\
&<16>&&<45>&&<10>&
\end{array}
\ee
where we have also displayed the minimal set of Higgs representations
needed to achieve the breaking pattern. The renormalization group
equations for this model are given by:

\bea
\frac{d \tilde\alpha}{d t} &=&  (\frac{7}{2} - 2n) \tilde\alpha^2 \nn
\frac{d Y_t}{d t} &=& (\frac{27}{4} \tilde\alpha - 14 Y_t ) Y_t
\eea
where n is the number of the extra copies of chiral superfields in
$(I + \bar{I})$ representations.
The results for the relevant cases are given in Table \ref{table:3} in
which it may be seen that the approach to the fixed point is slower
than in the other examples and the fixed point value of the top
coupling larger. For large n, however, the fixed point structure is
still likely to be important
in determining the low energy parameters.

\begin{table}\begin{center}
\begin{tabular}{|c|c|c|c|} \hline
$n$ & $(\frac{Y_t}{\tilde\alpha})^*$ & \EPP &
$(\frac{\tilde\alpha}{Y_t})_{SO(10)}^*/(\frac{\tilde\alpha}{Y_t})_{MSSM}^*$
\\ \hline\hline
0 & 0.23 & 0.91 & 1.70 \\ \hline
2 & 0.52 & 0.80 & 0.75 \\ \hline
4 & 0.80 & 0.69 & 0.49 \\ \hline
8 & 1.38 & 0.48 & 0.28 \\ \hline
\end{tabular}
\end{center}
\l{table:3}
\caption{SO(10)}
\end{table}

\subsection{Compactified (string) models}
As we mentioned above, in compactified theories the evolution of
couplings above the compactification scale is much faster (following a
power
law rather than a logarithmic evolution) leading to a very small value for
$(\frac{\alpha(M_C)}{\alpha(M_P)})^B$.
To illustrate this consider first the simple case
of a  dimension D=5 theory in which one dimension is a circle of
radius $R=\Lambda_c^{-1}$. We are interested in computing the one loop
corrections to the effective action, for example a loop with two
external gauge bosons at a scale Q. Then one finds for the polarisation
tensor the form \cite{vt}
\be
\Pi^{\mu\nu}=i(Q^{\mu}Q^{\nu}-Q^2g^{\mu\nu})\Pi(Q)
\ee
where
\be
\Pi(Q)\approx\beta_0\sum_{n=0}^{n=\Lambda_s R}\int
\frac{d^4P}{(2\pi^4)}
\frac{1}{P^2+n^2\Lambda_c^2}\frac{1}{(P+Q)^2+n^2\Lambda_c^2}
\l{eq:pir}
\ee
Here $\Lambda_c$ is the compactification scale and $\Lambda_s $ is the
scale beyond which the theory changes, for example the string scale,
and above which it makes no sense to use an effective field theory
approximation. The sum includes the Kaluza Klein modes with mass,
$n\Lambda_c$, less than or equal to $\Lambda_s$. One would think that
the appearance of such massive states would invalidate the use of mass
independent renormalisation group equations but this is not the case
for the sum of their contributions just serves to change the original
propagator in 4D to one in 5D leaving the original renormalisation
group equation intact. At scales $Q<<1/R$ evaluation of eq(\ref{eq:pir})
gives
\be
\Pi(Q)\approx\frac{\beta_0}{(4\pi)^2}(ln (QR)^2-2(\Lambda_sR-1))
\ee
The important point to note is that the integration over the range
$\Lambda_c<P<\Lambda_s$ generates a {\it power},
$(\Lambda_s/\Lambda_c)$ instead of $log(\Lambda_s/\Lambda_c)$. This
happens because in this range the theory
is effectively 5 dimensional and so the loop contribution is approximately
\bea
\Pi(Q)&\approx &i\beta_0\int_{\Lambda_c}^{\Lambda_s}
\frac{d^5P}{(2\pi^4)}
\frac{1}{P^2+n^2\Lambda_c^2}\frac{1}{(P+Q)^2+n^2\Lambda_c^2} \nn
&\approx&-2\frac{\beta_0}{(4\pi)^2}(\Lambda_sR-1)
\l {eq:ta}
\eea
As a result the effect of running in this region is enhanced
\be
\alpha^{-1}(t_c)=\alpha^{-1}(0)+\frac{b}{4\pi}\left(\frac{M_P}
{\Lambda_c}\right)^{D-4}
\l{eq:ar}
\ee
where $(D-4)$ is the number of compactified dimensions. Even for a
small difference between the compactification and Planck scale the
change from
logarithmic to power law evolution will make the factor
$(\frac{\alpha(M_C)}{\alpha(M_P)})^B$
very small implying that the infra red fixed point structure will
dominate the determination of couplings.

Of course the discussion so far has been much oversimplified for
compactification on a circle does not lead to realistic
theories. However the  effect persists, if somewhat ameliorated, in
realistic compactifications. This has been most thoroughly studied in
the context of superstring theories constructed via orbifold
compactification in which the terms we have been discussing are
known as threshold effects of the Kaluza Klein modes
\cite{l44,l445,at}. The simplest example, the symmetric orbifold
models, have Kaluza Klein modes which fall into N=4, N=2 and N=1
supermultiplets.
The former do not renormalise the gauge couplings but the latter two do giving
\be
\alpha^{-1}_a(M_s)=\alpha^{-1}(0)-\frac{\tilde b_a}{4\pi}
ln(|\eta (iT)|^4(T+\bar{T}))+c_a
\l {eq:te}
\ee
where $c_a$ are moduli  independent constants coming from the $N=1$
sector and  and $\tilde b_a$ is the beta function of the gauge group
factor in the N=2 sector\footnote{For states of modular weight -1 and
vanishing
Green Schwarz term $\tilde b =b$.}.
T is the expectation value of a moduli field which sets the scale of
compactification, the toroidal radius being given by $R^2=Re(T)$. The
function $\eta (T)$ is the Dedekind eta-function ($\eta
(T)q^{1/24}\Pi_n(1-q^n), \; q(T)=exp(-2\pi T)$) We see therefore that
the naive expectations are changed in orbifold compactification due to
the different multiplet structure of the massive excitations. The
effects do persist however, albeit somewhat reduced. To see this
explicitly let us expand eq(\r {eq:te}) for large T corresponding to
large radius of compactification\footnote{Note that in string theories
the string scale $\Lambda_s$ is somewhat larger than the
compactification scale $\Lambda_c$
by a factor $\le O(10)$}. This gives
\be
\alpha^{-1}_a(t_c)\approx\alpha^{-1}(0)+\frac{\tilde b_a}{4\pi}
\left(\frac{M_P}{\Lambda_c}\right)^2
\ee
It may be seen that, up to the Green Schwarz term and for modular
weight -1 fields, this is just the result of eq(\r {eq:ar}) with {\it
two}
compactified dimensions corresponding to the sector with N=2.

Of course the fixed point structure in general also changes above the
compactification scale. We have seen that the relevent $\beta$
function is that of the N=2 sector which in general may differ from
the theory below the compactification scale. The Yukawa coupling
evolution also changes. The wave function threshold corrections of an
untwisted field $A_j$ associated with the j-th of the
three internal compactified planes of the orbifold has been computed \cite{at}
\be
Y_j=2\tilde \gamma_j ln(|\eta (iT_j)|^4(T_j+\bar{T_j}))+y_j
\ee
where $y_j$ is a moduli-independent constant and the coefficient
$\tilde \gamma_j$ is the anomalous dimension of the $A_j$-field in the
corresponding $N=2$ supersymmetric theory. Since this field belongs to
an $N=2$ vector supermultiplet $\tilde\gamma_j=-\tilde b_j/2$ where
$\tilde b_j$ is the corresponding $\beta$ function coefficient of any
gauge subgroup that transforms $A_j$ non-trivially in the embedding
$N=2$ theory. The Yukawa coupling comes from the term in the
superpotential $W=2 A_1 A_2 A_3$. Including the wave function
normalisations the tree level physical Yukawa coupling is
$\lambda_{123}=\sqrt{2}g$ the relation to the gauge coupling
corresponding to the fact the fields belong to N=2
supermultiplets. Including the effects of string threshold corrections
(i.e. the Kaluza Klein modes) gives
\be
Y_{123}(M_s)=\frac{2\tilde\alpha_a}{1+4\pi\alpha_a(M_s)(y_1+y_2+y_3)}
\ee
After a little algebra this may be rewritten in the form
of eq(\ref{eq:soln1}) with
\bea
t&=&(|\eta (iT_j)|^4(T_j+\bar{T_j})) \nn
B&=& 1\nn
\left(\frac{Y_{123}}{\tilde \alpha}\right)^*_{orbifold}&=&2
\l{eq:compact}
\eea
The fixed point corresponds to the N=2 value $\lambda= \sqrt{2}g$ when
the initial N=2 breaking terms $y_i$ become irrelevant. The rate of
approach is just determined by $\alpha(t)/\alpha(0)$. Due to the power
law evolution this may be very small in theories with a lot of matter
and a non-asymptotically free coupling in the $N=2$ sector. For
example with the same $\beta$ function as in the $SU(3)^3$ case,
normalising the coupling at $t_s$ to be the unified coupling 1/24, we
would get the tree level coupling, $\tilde \alpha(0)$  to be of O(1)
for $M_c/M_s$=1/5! In this case a reasonable estimate
for ($\frac{\alpha(M_X)}{\alpha(M_S)})^B$  is 1/24 implying the
fixed point is very closely approached.

Of course the value of the couplings at $M_s$ must be fed into the RG
equations for the theory below the compactification scale to take it
to $M_X$ where the MSSM RG equations take over. To quantify this we
simply take the focusing formula from eq(\ref{eq:focus}). Using
eq(\ref{eq:compact}) shows $x(M_S)\approx 24 x(0)$ so the range of
values in the domain of attraction of the MSSM
fixed point is very large. If $x(M_s) \leq 1/2$ then after
focusing the range of values becomes $ 209 \leq m_t  \leq 210$.

\section{Summary}
In conclusion we have found that the infra-red fixed point structure
of the unified theory beyond the \sm is likely to play a very
important and in many cases a dominant role in the determination of
ratios of the Yukawa to gauge couplings. This means that these
couplings may be determined simply from a knowlege of the multiplet
content and gauge group structure without needing to know their value
in the underlying ``Theory of Everything". (In string theories this
may avoids the  difficult question of determining the moduli
dependence of the couplings and the values of the moduli). The most
obvious prediction is for the top quark mass and we have argued that
it is very likely to be very close to  its quasi fixed point value.
However many further predictions for the effective low-energy theory
follow from the IR fixed point structure corresponding to the
appearance of the hidden symmetries of the RG equations. For example
unrelated Yukawa couplings will be driven to be equal at the fixed point
if they involve chiral superfields carrying the {\it same}
 gauge quantum numbers as in our $SU(3)^3$ example
discussed above, the RG fixed point structure will make them equal.
Similar conclusions follow for other fixed point structures not
discussed here relating the soft SUSY breaking  mass terms of the MSSM
\cite{inpreparation}. It may therefore be hoped that a study of this
type of Infra Red fixed points will shed light on the values of all
the parameters of the \sm and not just the top quark mass
predictions presented here.

{\bf Acknowledgments} One of us (GGR) would like to thank C.Kounnas,
G.Veneziano and T.Taylor for discussions and G.Kane for a conversation
which led to this return visit.

\end{document}